\documentclass[10pt,doublecolumn,twoside]{IEEEtran}

\usepackage{etoolbox}
\newtoggle{doublecolumn}
\newtoggle{calculateFigures}

\toggletrue{doublecolumn}
\togglefalse{calculateFigures}
\usepackage{etex}

\usepackage[utf8]{inputenc}
\usepackage[T1]{fontenc}

\usepackage[english]{babel}
\usepackage{lipsum}
\usepackage{amsbsy,amsmath,amsfonts,amssymb,amsthm}
\usepackage{mathtools}
\usepackage{textcomp} %degrees centigrade symbols, euros, etc. 
\usepackage{relsize}

\usepackage{bm,cite,cases,pstricks,times,url,verbatim} % try to remove this line for http://www.conf-express.org
\usepackage[noend]{algpseudocode}
\usepackage{booktabs}
\usepackage{tabularx,array,dcolumn,multirow}

\usepackage{algorithm}

\usepackage{soul} %\ul per underline

\usepackage{blkarray,bigdelim} % matrices with braces

\allowdisplaybreaks[4]

\usepackage{centernot} % prints the sym­bol \not on the fol­low­ing ar­gu­ment

\newcommand{\subparagraph}{}

\iftoggle{doublecolumn}{
    \newtheorem{thm}{Theorem}
    \newtheorem{fact}{Fact}
    \newtheorem{lemma}{Lemma}
    \newtheorem{definition}{Definition}
    \newtheorem{conj}{Conjecture}
    \newtheorem{propos}{Proposition}
    \newtheorem{corol}{Corollary}
    \newtheorem{ass}{Assumption}
    \newtheorem{example}{Example}
    \newtheorem{remark}{Remark}
    \newtheorem{note}{Note}
    \newtheorem{obs}{Observation}

}{

    \newtheoremstyle{exampstyle}
      {0} % Space above
      {0} % Space below
      {\itshape} % Body font
      {} % Indent amount
      {\bfseries} % Theorem head font
      {.} % Punctuation after theorem head
      {.5em} % Space after theorem head
      {} % Theorem head spec (can be left empty, meaning `normal')

    \theoremstyle{exampstyle} \newtheorem{thm}{Theorem}
    \theoremstyle{exampstyle} 
    \theoremstyle{exampstyle} \newtheorem{lemma}{Lemma}
    \theoremstyle{exampstyle} 
    \theoremstyle{exampstyle} 
    \theoremstyle{exampstyle} 
    \theoremstyle{exampstyle} \newtheorem{corol}{Corollary}
    \theoremstyle{exampstyle} \newtheorem{ass}{Assumption}
    \theoremstyle{exampstyle} 
    \theoremstyle{exampstyle} 
    \theoremstyle{exampstyle} 
    \theoremstyle{exampstyle} 
}

\newcommand{\argmin}[1]{\operatorname{arg}\,\underset{#1}{\operatorname{min}}\;}

\makeatletter
\newcommand{\pushright}[1]{\ifmeasuring@#1\else\omit\hfill$\displaystyle#1$\fi\ignorespaces}
\newcommand{\pushleft}[1]{\ifmeasuring@#1\else\omit$\displaystyle#1$\hfill\fi\ignorespaces}

\begingroup
\catcode`\#=11
\gdef\noautorotate{-dAutoRotatePages#/None}
\endgroup

\usepackage{graphicx,xcolor,float,dblfloatfix}
\usepackage{psfrag}

\usepackage{caption}
\usepackage{subcaption}

\newcommand{\subalign}[1]{%
  \vcenter{%
    \Let@ \restore@math@cr \default@tag
    \baselineskip\fontdimen10 \scriptfont\tw@
    \advance\baselineskip\fontdimen12 \scriptfont\tw@
    \lineskip\thr@@\fontdimen8 \scriptfont\thr@@
    \lineskiplimit\lineskip
    \ialign{\hfil$\m@th\textstyle##$&$\m@th\textstyle{}##$\crcr
      #1\crcr
    }%
  }
}

\iftoggle{calculateFigures}{
    \usepackage[crop=off,runs=2,pspdf=\noautorotate]{auto-pst-pdf}
}{
    \usepackage[off]{auto-pst-pdf}
}

\usepackage{hyperref} %[hidelinks]

\graphicspath{%
{./Figures/}
}

\newcommand\defeq{\stackrel{\mathclap{{\rm def}}}{=}}

\usepackage{color}

\usepackage{caption}
\usepackage{subcaption}
\usepackage[compact]{titlesec}
\begin{document}

\title{Joint Optimization of Energy Efficiency \\ and Data Compression in TDMA-Based \\Medium Access Control for the IoT}
%\title{An Analysis \\of the Energy/Distortion Tradeoff in the IoT}
%\title{On the Energy/Distortion Tradeoff in the IoT}

\author{\IEEEauthorblockN{Chiara~Pielli, Alessandro~Biason, Andrea~Zanella and~Michele~Zorzi}\\
\IEEEauthorblockA{\{piellich,biasonal,zanella,zorzi\}@dei.unipd.it\\
Department of Information Engineering, University of Padova - via Gradenigo
6b, 35131 Padova, Italy
}%
}

\maketitle
\pagestyle{empty}
\thispagestyle{empty}

%\fontdimen2\font=3.1pt

\begin{abstract}
Energy efficiency is a key requirement for the Internet of Things, as many sensors are expected to be completely stand-alone and able to run for years without battery replacement. Data compression aims at saving some energy by reducing the volume of data sent over the network, but also affects the quality of the received information. In this work, we formulate an optimization problem to jointly design the source coding and transmission strategies for time-varying channels and sources, with the twofold goal of extending the network lifetime and granting low distortion levels. We propose a scalable offline optimal policy that allocates both energy and transmission parameters (i.e., times and powers) in a network with a dynamic Time Division Multiple Access (TDMA)-based access scheme.
\end{abstract}

\section{Introduction}\label{sec:introduction}

In the Internet of Things (IoT) a large number of heterogeneous devices are expected to exchange data gathered from the surrounding environment.
Although many IoT devices may be connected to the energy grid all the time (e.g., in smart house applications), most of them will have to rely on their own limited energy supply and will likely be deployed in remote or harsh places~\cite{lazarou}. The burden of replacing sensors or recharging their batteries every few weeks may outweigh all the benefits of collecting data, and nodes failure due to power depletion may even lead to the breakdown of the whole architecture~\cite{liang}. Thus, it is crucial to conserve as much energy as possible. %, within the limits imposed by the Quality of Service (QoS) requirements.
On the other hand, periodic sampling of environmental signals leads to enormous amounts of raw data, the transmission of which would rapidly deplete the sensor energy. One of the key strategies to solve this problem is \emph{data compression}, which allows to reduce the amount of transmitted data while maintaining high levels of Quality of Service (QoS).
The goal of this work is to investigate the trade-offs between energy consumption and data compression at the Medium Access Control (MAC) layer.

The energy efficiency problem has gained much interest in the last years and several protocols have been proposed with the target of extending the network lifetime as much as possible. A lot of effort has been put into the design of the MAC layer~\cite{hac}, since the usage of the RF chain may have a major impact on the energy consumption.
Many works in the literature study both offline and online policies for the energy allocation problem~\cite{cassandras}, often in the presence of Energy Harvesting (EH), but they usually focus on a single transmitter-receiver pair~\cite{ho, fu, Biason2015d}.

Also the idea of QoS provisioning at the MAC layer is not new, but the QoS metrics considered are typically throughput, latency, and delivery ratio~\cite{natkaniec, yigitel} and much fewer works take into account the effects that signal processing has on the transmitted information. Indeed, although compression allows for some energy savings, due to the reduced number of symbols to be sent, it affects the received information by introducing a certain degree of distortion.
In~\cite{dey}, the authors study energy allocation policies to minimize the signal distortion when several sensors measure the same process of interest and exploit data fusion techniques, whereas in~\cite{tapparello} an online joint source coding and data transmission optimization strategy is investigated for sensors with EH capabilities that generate correlated information.

 %In~\cite{yang}, the authors propose an optimal packet scheduling policy for two EH transmitters that leverages on an iterative backward waterfilling algorithm.

Often, in the literature, uncoordinated access schemes are chosen because of their flexibility and lower synchronization costs. Nevertheless, coordinated access schemes completely avoid collisions and interference since there is no channel contention and, by adopting appropriate duty cycling mechanisms, also the energy wastage due to idle listening is prevented. %Sensor-MAC (SMAC)~\cite{smac} has been the first protocol to introduce the idea of periodic synchronous active and inactive periods, and organizes the newtork into virtual clusters of nodes that follow the same duty cycling schedule and that communicate between each other via border nodes that follow more than one schedule. Major drawbacks of S-MAC are the rigidity introduced by the use of active periods of fixed size and the delay caused by sleeping.
%A well-known TDMA-based algorithm is PEDAMACS~\cite{Cionca,5}, which employs both centralized and distributed graph node coloring to find the shortest schedule possible and thus minimize latency. 
%Many protocols combine TDMA and Channel Sensing Multiple Access (CSMA) techniques~\cite{trama, T-MAC, ZMAC}, or exploit the information from the neighbouring nodes to coordinate the use of the time slots,~\cite{incel, vanHoesel}. 
Recently, the Internet Engineering Task Force introduced the Time-Slotted Channel Hopping (TSCH)~\cite{tsch} mode as an amendment to the MAC portion of the IEEE802.15.4e standard. % and it now represents the major emerging standard for industrial automation and process control Low-power and Lossy Networks (LLNs). 
It is a TDMA-based scheme and adopts a channel-hopping mechanism to improve reliability in the presence of narrowband interference and multi-path fading. The standardization of 6TiSCH (IPv6 over TSCH) enabled a wide-spread use of the TSCH mode in industrial networks, leading to a revival of TDMA-based schemes.

In this paper we propose a synchronized MAC protocol for an IoT network with an arbitrary number of nodes. We develop a TDMA-like scheme based on an optimization framework, which adopts convex and alternate programming to minimize the data distortion and extend the network lifetime simultaneously, under QoS constraints. Realistic energy consumption models that consider both the compression and transmission costs are taken into account.

The paper is organized as follows. In Section~\ref{sec:system_model} we describe our system model and introduce the optimization problem, which is divided in two parts, namely the Frame-Oriented Problem and the Energy-Allocation Problem, which are solved in Sections~\ref{sec:FOP} and~\ref{sec:EAP}, respectively. Section~\ref{sec:numerical_evaluation} shows the numerical evaluation. Finally, Section~\ref{sec:conclusions} concludes the paper.

\emph{Notation:} Boldface letters are used for matrices and vectors; $\mathbf{E}_i$ refers to the $i$-th row of matrix $\mathbf{E}$, $\mathbf{E}^{(k)}$ to the $k$-th column, and $E_i^{(k)}$ is the $(i,k)$ element. With ``$\forall i$'' and ``$\forall k$'', we summarize $i = 1,\ldots,N$ and $k = 1,\ldots,n$, respectively, where $N$ and $n$ are defined in the next section.

%\clearpage
\section{System model}\label{sec:system_model}

We consider a network of $N$ users which access the uplink channel to send data packets to a central Base Station (BS) using TDMA.
Time is divided into frames and frame $k$ corresponds to the time interval $[t_k,t_{k+1})$.

We consider the channel gains to be constant during each frame, and we approximate the average physical rate of user $i\in\mathcal{N} \defeq \{1,\dots,N\}$ as:~
\begin{align}
    r_i^{(k)} = W \log_2 \left( 1+\gamma_i^{(k)} \right) = W \log_2 \left( 1+h_i^{(k)} P_{{\rm tx},i}^{(k)}\right),   \label{eq:rate}
\end{align}
where $W$ is the bandwidth, $\gamma_i^{(k)}$ the average Signal-to-Noise Ratio (SNR) of user $i$ in frame $k$, $P_{{\rm tx},i}$ the transmission power used by the node, and $h_i^{(k)}$ the average channel gain normalized with respect to noise.

As in~\cite{ulukus}, we adopt an information-theoretic approach in which full Channel State Information (CSI) is available a priori. This allows us to derive the optimal policy to minimize the average distortion over time and to find useful bounds on the actual performance that can be obtained in practice.

\subsection{Data Generation and Compression}\label{subsec:data_model}
Nodes may generate data by collecting measurements from the environment or by serving as relays to the central BS for farther nodes. %The data collection rate of node $i$ during frame $k$ is $\lambda_i^{(k)}$ and is assumed to be constant during the frame period. 
Before transmission and according to the type of signal generated, nodes can perform compression in order to limit the physical amount of data to be sent over the network. The distortion degree $D_i^{(k)}$ is a function of the compression ratio, $\eta_{C,i}^{(k)} =  L_{i}^{(k)}/L_{0,i}^{(k)}$, where  $L_i^{(k)}$ is the size of the compressed packet, and $L_{0,i}^{(k)}$ is the size of the original data. We define the following mathematical expression, which approximates the rate-distortion curve of a Gaussian source~\cite{berger}:~
\begin{align}
    D_i^{(k)} =  b_i \left(\dfrac{1}{(\eta_{C,i}^{(k)})\,^{\alpha_i}} - 1\right), %=b_i^{(k)} \left( {\left(\dfrac{L_{0,i}^{(k)}}{L_i^{(k)}}\right)}^{\alpha_i^{(k)}} - 1\right) 
    \label{eq:dist}
\end{align}
  
\noindent where $\alpha_i, b_i > 0$. Notice that the distortion is null when the packet is not compressed, i.e., $\eta_{C,i}^{(k)}=1$.

We also introduce a QoS requirement on the quality of the received data $D_i^{(k)} \leq D_{{\rm th}, i}^{(k)}$, where $D_{{\rm th}, i}^{(k)}$ is a threshold distortion level: if the reconstruction error
exceeds this threshold, the signal generated by the source node is no longer useful for the final destination. %This limit may be imposed, e.g., by the application on top of the system or by the impossibility to extract useful information from the received packet if the distortion level were too high.

\subsection{Energy Consumption Model}\label{subsec:energy_model}

Devices are battery-equipped and the battery level of node $i$ in frame $k$ is $B_i^{(k)}$. Since no harvesting sources are considered, the initial battery levels $\mathbf{B}^{(0)}$ will strongly impact the system performance. In every frame, an energy $E_i^{(k)}\in [0, B_i^{(k)}]$ is used according to the following sources of energy consumption.

\emph{\textbf{Data processing}} the energy spent to process the data gathered from the environment or other nodes can be characterized by exploiting the results of~\cite{zordan}:~ 
 \begin{align}
   E_{P,i}^{(k)}  = \hat{E}_{0,i} \cdot L_{0,i}^{(k)} \cdot N_{C,i}(\eta_{C,i}^{(k)}),    \label{eq:e_processing1}
 \end{align}

 \noindent where $\hat{E}_{0,i}$ is the energy consumption per CPU cycle which thus depends on the processor of the node, and $ N_{C,i}(\eta_{C,i}^{(k)})$ is the number of clock cycles per bit needed to compress the input signal and is a function of the compression ratio. $N_{C,i}(\eta_{C,i}^{(k)})$ depends on the compression algorithm (see~\cite{zordan} for further details), and, in the case of Lightweight Temporal Compression (LTC) and Fourier-based Low Pass Filter (DCT-LPF), it is linear: $N_{C,i}(\eta_{C,i}^{(k)}) = 
 \hat{\alpha_i}\eta_{C,i}^{(k)}  + \hat{\beta_i}$.
We assume the devices use one of these algorithms, so the energy consumption due to processing becomes:~
 \begin{align}
   E_{P,i}^{(k)}  = \hat{E}_{0,i} L_{0,i}^{(k)} \left(\hat{\alpha_i} \frac{L_{i}^{(k)}}{L_{0,i}^{(k)}}  + \hat{\beta_i}\right) = E_{0,i}  L_i^{(k)} + \beta_{P,i}^{(k)},   \label{eq:e_processing}
 \end{align}
 
 \noindent where we defined $E_{0,i} \triangleq \hat{E}_{0,i}\,\hat{\alpha_i }$ and $\beta_{P,i}^{(k)} \triangleq  \hat{E}_{0,i}\,\hat{\beta_i}\,L_{0,i}^{(k)}$.

\emph{\textbf{Data transmission}} the energy spent for the data transmission task can be expressed as:~
\begin{align}
 E_{TX,i}^{(k)} =  P_{{\rm tx},i}^{(k)} \cdot \tau_i^{(k)},   \label{eq:e_tx}
\end{align}

  \noindent where $\tau_i^{(k)}$ is the transmission duration and $P_{{\rm tx},i}^{(k)}$ is the transmission power, which is assumed to be constant during the whole transmission. %and to be lower and upper bounded by $P_{i,\rm min}^{(k)}$ and $P_{i,\rm max}^{(k)}$, respectively.

\emph{\textbf{Data sensing and circuitry costs}}
we also consider the contributions to the energy consumption of both sensing operations and energy losses due to circuitry, which include, e.g., the energy spent for node switches from sleep mode to active mode and viceversa, the synchronization costs, and the additional energy lost during the transmission. 
We can express these quantities in the following way:~
\begin{align}
 E_{C,i}^{(k)} =  \beta_{{\rm sens},i}^{(k)} + \beta_{C,i}^{(k)} + \mathcal{E}_{C,i} \cdot \tau_i^{(k)},  \label{eq:e_circuitry}
\end{align}

 \noindent where $\beta_{{\rm sens},i}^{(k)}$ and $\beta_{C,i}^{(k)}$ represent the constant sensing and circuitry contributions, respectively, and $\mathcal{E}_{C,i}$ is the rate of circuitry energy consumption during data transmission.
Note that the energy consumption due to collisions and overhearing is avoided because of the exclusive use of the communication channel guaranteed by our TDMA approach. %TODO synchronization among nodes in TDMA: magari riusciamo a includerla nella formula precedente

By combining Eqs.~\eqref{eq:e_processing}-\eqref{eq:e_circuitry}, the total energy consumption of a node in a single frame $k$ is:~
\begin{align} \label{eq:energy_used}
\begin{split}
  E_{{\rm used},i}^{(k)}  & =E_{P,i}^{(k)} + E_{TX,i}^{(k)} + E_{C,i}^{(k)} \\
  & = \beta_{i}^{(k)} + E_{0,i} L_i^{(k)} + (P_{{\rm tx},i}^{(k)} + \mathcal{E}_{C,i}^{(k)} ) \tau_i^{(k)},
 \end{split}
\end{align}

 \noindent where $\beta_i^{(k)} \triangleq \beta_{P,i}^{(k)} + \beta_{{\rm sens},i}^{(k)} + \beta_{C,i}^{(k)}$.

\subsection{Optimization Problem} \label{subsec:opt_problem}

The goal of the system is to simultaneously satisfy the QoS requirements and extend the network lifetime. To handle these two conflicting objectives, we set up the following weighted optimization problem:~
\begin{align}
    \min_{\textstyle \{\mathbf{E}^{(0)},\mathbf{E}^{(1)},\ldots\}} \sigma \frac{1}{n} \sum_{k = 1}^n f_{FOP}^{(k)}(\mathbf{E}^{(k)}) - (1-\sigma) n, \label{eq:general_output}
\end{align}

\noindent where $\sigma$ is the weight in $[0,1]$ and $n$ is the effective lifetime of the system, which we defined as the first frame in which at least one node dies, i.e., it does not have enough energy in its battery to transmit any more data while satisfying the distortion constraint. Since the network lifetime is an outcome of the energy assignment, we consider $k \in \mathbb{N}$ for the optimization variables, i.e., the number of optimization variables is not known a priori. The second term of~\eqref{eq:general_output} is a decreasing function of $n$, and is used to express the trade-off between distortion and lifetime.

Given the energy consumption vector $\mathbf{E}^{(k)}, \, \forall k$, the lifetime $n$ is uniquely determined, whereas the distortion depends on the parameters $\bm{\tau}$, $\mathbf{L}$, and $\mathbf{P}_{\rm tx}$. With the final goal of minimizing the normalized distortion (the normalization is done with respect to the QoS threshold), we define function $f_{FOP}^{(k)}(\mathbf{E}^{(k)})$ as~
 \begin{align}
  f_{FOP}^{(k)}(\mathbf{E}^{(k)}) = \min_{\textstyle \bm{\tau}^{(k)}, \mathbf{L}^{(k)}, \mathbf{P}_{\rm tx}^{(k)}} \, \max_{i \in \mathcal{N}} \frac{D_i^{(k)}}{D_{{\rm th}, i}^{(k)}},
  \label{eq:FOP_minmax}
 \end{align}
 
\noindent which will be presented in its extended form in~\eqref{prob:FOP}.

We have structured the problem in a modular fashion that introduces a level of independence between the building blocks. These can be slightly adapted to meet different requirements in a separate way while keeping the overall framework.
The two blocks have the following objectives. 
\begin{enumerate}
 \item \emph{Energy Allocation Problem} (EAP): it is the main problem, with the goal expressed in Eq.~\eqref{eq:general_output}.  EAP defines the energy allocation over time $\{\mathbf{E}^{(0)},\mathbf{E}^{(1)},\ldots\}$;
 \item \emph{Frame-Oriented Problem} (FOP): the focus is on single frames and the goal of this sub-problem is to determine the transmission durations and powers that minimize Eq.~\eqref{eq:FOP_minmax} \emph{given} the energy consumed in that slot.
%\item \emph{Delay-Oriented Problem} (DOP): it aims at reducing latency within one frame when the transmission durations are given by FOP. Describing DOP is beyond the scope of this work.
\end{enumerate}

In practice, the two problems are tightly coupled: EAP defines the energy allocation to use in every slot, which is used by FOP to determine~\eqref{eq:FOP_minmax}; on the other hand, the output of FOP affects the choice of EAP (see~\eqref{eq:general_output}). In the next two sections, we discuss these two problems and how they are interrelated.

\section{Frame-Oriented Problem}\label{sec:FOP}

In this section, we present FOP, which defines the transmission durations and powers in a specific frame when the energy consumption is given. According to~\eqref{eq:FOP_minmax}, FOP envisages a conservative approach and aims at minimizing the maximum normalized distortion among all users.

\subsection{Optimization Problem}\label{subsec:FOP_optimization} 

The amount of energy consumed in a single frame, namely $E_i^{(k)}$, is given for all users and its optimal allocation is determined by EAP, which we will discuss in Section~\ref{sec:EAP}.

Since FOP addresses a single frame, for ease of notation we will omit the dependence on the time index $k$ throughout this section. Accordingly, boldface letters refer to column vectors that span over the $N$ users for the considered frame. The optimization problem we set up and solve is the following:~
\begin{subequations} \label{prob:FOP}
 \begin{flalign}
    \text{FOP:} && & \min_{\textstyle \bm{\tau}, \mathbf{L}, \mathbf{P}_{\rm tx}} \, \max_{i \in \mathcal{N}} \frac{D_i}{D_{{\rm th}, i}}, &
 \end{flalign}
 \vspace{-\belowdisplayskip}
 \vspace{-\abovedisplayskip}
 \begin{alignat}{2}
    \shortintertext{subject to}
    & D_i =  b_i \left( {\left(\dfrac{L_{0,i}}{L_i}\right)}^{\alpha_i} - 1 \right) \leq D_{{\rm th}, i}, \quad && \forall i, \label{eq:FOP_distortion}\\
    & L_i \leq \tau_i \, r_i, \quad && \forall i, \label{eq:FOP_capacity_const} \\
    & E_{0,i} L_i  + \beta_i + (P_{{\rm tx},i} + \mathcal{E}_{C,i}) \tau_i \leq E_i, \quad && \forall i, \label{eq:FOP_energy_const}\\
    & P_{{\rm min},i} \le P_{{\rm tx},i} \le P_{{\rm max},i}, \quad && \forall i, \label{eq:FOP_power}\\
    & 0 \le L_i \le L_{0,i}, \quad && \forall i,  \label{eq:FOP_L}\\
    &  \sum_{i = 1}^N \tau_i \le T. \label{eq:FOP_tau}
 \end{alignat}
\end{subequations}

\noindent The objective function represents a $\min\max$ problem in order to guarantee fairness among the users in the network. Constraint~\eqref{eq:FOP_distortion} represents the distortion as defined in~\eqref{eq:dist}.
Inequality~\eqref{eq:FOP_capacity_const} is a capacity constraint derived from the Shannon-Hartley theorem (see~\eqref{eq:rate}). However, since larger data rates lead to smaller transmission times for a given data size, the previous constraint can be taken with equality. By doing so, $ \mathbf{L}$ can be removed from the optimization variables and expressed as a function of $\mathbf{P}_{\rm tx}$ and $\bm{\tau}$.  Inequality~\eqref{eq:FOP_energy_const} represents the relation between the given energy $E_i$ and the consumed one (see~\eqref{eq:energy_used}). Without loss of generality, we set Constraint~\eqref{eq:FOP_energy_const} with equality, as otherwise a positive amount of energy would be wasted.
%Notice that when~\eqref{eq:FOP_energy_const} is satisfied with strict inequality, a certain amount of energy is wasted.
%however, imposing the equality may not necessarily lead to an optimal solution. 
%indeed, with higher amounts of effectively consumed energy, the transmission durations increase, and satisfying Constraint~\eqref{eq:FOP_tau} becomes more difficult. 
%Since it is responsibility of EAP to determine the optimal amount of energy each sensor can use in the considered frame, we set Constraint~\eqref{eq:FOP_energy_const} with equality, without loss of generality.
 We also set the realistic bounds $P_{{\rm min},i}$ and $P_{{\rm max},i}$ on the transmission power to reflect the physical transmission capabilities of a device.\footnote{Formally, we should also consider the case $P_{{\rm tx},i} = 0$, but this would lead to an arbitrarily large distortion, which does not represent a relevant case for our study.} Constraint~\eqref{eq:FOP_L} imposes that the number of transmitted bits $L_i$ is not larger than the number of generated bits $L_{0,i}$. Finally, the last constraint combines together all the users in a TDMA fashion. Note that, without~\eqref{eq:FOP_tau}, FOP could be decomposed into $N$ separate problems.

\subsection{Solution of \emph{FOP}} \label{subsec:FOP_sol}

We now describe how to solve FOP. First, note that the objective function can be formulated in an equivalent way by introducing an auxiliary optimization variable $\Gamma$ as follows:~
\begin{subequations} \label{prob:FOP_gamma}
  \begin{flalign}
      \text{FOP$_\Gamma$:} && & \min_{\textstyle \Gamma, \bm{\tau}, \mathbf{P}_{\rm tx}} \Gamma,  & \label{eq:FOP_objective_gamma}
  \end{flalign}
  \vspace{-\belowdisplayskip}
  \vspace{-\abovedisplayskip}
  \begin{alignat}{2}
  \shortintertext{subject to}
  & \frac{D_i}{D_{{\rm th}, i}} \leq \Gamma, \quad  \forall i, \label{eq:FOP_gamma} \\
  & \mbox{Constraints } \eqref{eq:FOP_distortion}-\eqref{eq:FOP_tau}.
  \end{alignat}
\end{subequations}

The optimal solution of \text{FOP$_\Gamma$}, namely $\Gamma^\star$, depends on the energy allocated to each user by EAP in the considered frame since the optimization variables are interrelated through Eq.~\eqref{eq:FOP_energy_const}, and thus we express it as $\Gamma^\star = f_{\rm FOP}(\mathbf{E})$. Notice that we are only interested in solutions $\Gamma^\star\le 1$ (i.e., every distortion is below threshold), as will be explained in Section~\ref{subsec:FOP_unfeasible}.
We now propose an efficient technique to extract one optimal solution.

\begin{lemma}\label{lemma:FOP_gamma}
    In at least one optimal solution, Constraint~\eqref{eq:FOP_gamma} is satisfied with equality for every $i$.
    \begin{proof}
        See \appendixname~\ref{proof:FOP_gamma}.
    \end{proof}
\end{lemma}

Throughout this subsection, consider a fixed $\Gamma$.
First of all, notice that, the higher $P_{{\rm tx},i}$, the shorter $\tau_i$, according to~\eqref{eq:FOP_capacity_const}. Therefore, with higher transmission powers, it is more likely to satisfy the constraint on the frame duration~\eqref{eq:FOP_tau}. Based on this consideration, we choose the highest $P_{{\rm tx},i}$ that satisfies both~\eqref{eq:FOP_energy_const} and~\eqref{eq:FOP_power}. Hence, if we combine~\eqref{eq:FOP_capacity_const} taken with equality and~\eqref{eq:FOP_energy_const}, we obtain:~
\begin{align}\label{eq:P_TX_E}
     g_i(P_{{\rm tx},i}) \leq \frac{W}{L_i}(E_i - E_{0,i} L_i  + \beta_i),
\end{align}

\begin{figure}[t]
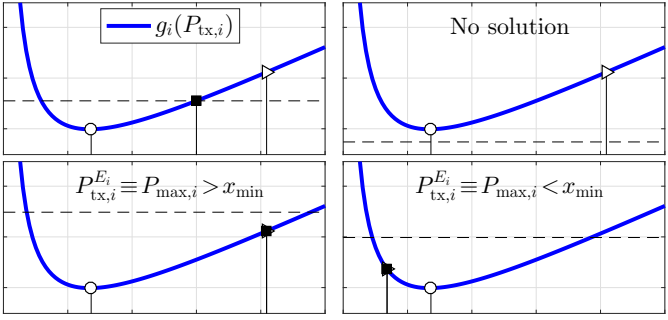

    \centering
    \begin{minipage}[t]{.49\columnwidth}
        \centering
        \includegraphics[trim = 0mm 0mm 0mm 0mm,  clip, width=1\columnwidth]{plot_g1.eps}
    \end{minipage}%
    \hfill%
    \begin{minipage}[t]{.49\columnwidth}
        \centering
        \includegraphics[trim = 0mm 0mm 0mm 0mm,  clip, width=1\columnwidth]{plot_g2.eps}
    \end{minipage}\\
    \begin{minipage}[t]{.49\columnwidth}
        \centering
        \includegraphics[trim = 0mm 0mm 0mm 0mm,  clip, width=1\columnwidth]{plot_g3.eps}
    \end{minipage}%
    \hfill%
    \begin{minipage}[t]{.49\columnwidth}
        \centering
        \includegraphics[trim = 0mm 0mm 0mm 0mm,  clip, width=1\columnwidth]{plot_g4.eps}
    \end{minipage}\\
    \hspace{.005\textwidth}
    \begin{minipage}[t]{1\columnwidth}
        \caption{Function $g_i(P_{{\rm tx},i})$. The dash-line represents different values of $W/L_i(E_i - E_{0,i} L_i  + \beta_i)$. The empty circle and triangle markers represent $x_{\rm min}$ and $P_{{\rm max},i}$, respectively. The black square markers represent $P_{{\rm tx},i}^{E_i}$.}
        \label{fig:plot_g}
    \end{minipage}
    \vspace{-.5cm}
\end{figure}

\noindent where we defined $g_i(x) \triangleq (x + \mathcal{E}_{C,i})/\log_2(1+h_i\,x)$ (see \figurename~\ref{fig:plot_g}).
Note that all the terms on the right-hand side are fixed since $E_i$ is given, $L_i$ is derived from $\Gamma$ through Lemma~\ref{lemma:FOP_gamma} and the remaining are system parameters. It can be shown that $g_i(x)$ is a decreasing-increasing function of~$x$ and that it admits only one minimum. We now make the following technical assumption.
\begin{ass} \label{ass:P_max_high}
    $x_{\rm min} \defeq \argmin{x} \{g_i(x)\} < P_{{\rm max},i},\ \forall i$.
\end{ass}

When the previous assumption is not satisfied, we choose $P_{{\rm max},i}$ as transmission power.
A method to find the optimal $\Gamma^\star$ is then based on the following reasoning. Let $P_{{\rm tx},i}^{E_i}$ be the maximum power at which~\eqref{eq:P_TX_E} is satisfied with equality. Then, the optimal $P_{{\rm tx},i}$ will be\footnote{ We recall that choosing higher $P_{{\rm tx},i}$ leads to lower $\tau_i$}~
\begin{align}\label{eq:P_TX_star}
    P_{{\rm tx},i}^\star = \min\{P_{{\rm max},i},P_{{\rm tx},i}^{E_i}\}.
\end{align}

\noindent If $P_{{\rm tx},i}^{E_i}$ does not exist or $P_{{\rm tx},i}^{E_i} < P_{{\rm min},\ell}$, then no solution exists and the problem is infeasible for the given $\Gamma$. %However, notice that the converse does not hold, since we are neglecting the TDMA constraint.
%TODO queste frasi sono un po' confuse, da sistemare magandi dando un esempio grafico nella versione estesa

We now present a key result that will be used to solve FOP.

\begin{thm}\label{thm:Gamma1_Gamma2}
    If \emph{FOP}$_\Gamma$ is feasible for a fixed $\Gamma^\prime \le 1$, then \emph{FOP}$_\Gamma$ is feasible for all $\Gamma^{\prime\prime}$ such that $\Gamma^\prime \leq \Gamma^{\prime\prime} \leq 1$.
    \begin{proof}
        See \appendixname~\ref{proof:Gamma1_Gamma2}.
    \end{proof}
\end{thm}

Theorem~\ref{thm:Gamma1_Gamma2} implies the following corollary.
\begin{corol}
    $\Gamma^\star$ can be found with a bisection search over the interval $[0, 1]$.
\end{corol}

%Therefore, for every user $i$, FOP can be solved with a bisection search over $\Gamma$, 
Thus few operations are required to find $\Gamma^\star$ with a very high precision. When $\Gamma^\star$ has been found, all the other parameters can be found using Lemma~\ref{lemma:FOP_gamma} and Eq.~\eqref{eq:P_TX_star}.

\subsection{Infeasibility of the Problem}\label{subsec:FOP_unfeasible}
 
We consider FOP to be infeasible if either of these two conditions occurs: i1) at least one constraint is not satisfied, i.e., there exists no allocation of $\bm{\tau}$ and $\mathbf{P}_{\rm tx}$ that allows all users to transmit their packets within the frame duration and with the assigned energy levels, i2) the constraints are satisfied but  $\Gamma^\star > 1$, i.e., $\exists i\in\mathcal{N} : D_i>D_{{\rm th}, i}$, which can be interpreted as a violation of the QoS constraint.

% \AB{stringerei molto questa sezione. Quale frase possiamo aggiungere qua?}

Since we consider hard constraints, if FOP were infeasible, the only strategies available would be: 1) allocating a larger amount of energy, 2) choosing a longer frame duration, or 3) removing some users. In this paper, we study technique 1) in Section~\ref{sec:EAP}, where we solve the energy allocation problem, and leave 2) and 3) as part of our future work. 

%We observe, however, that 2) is more suitable when a lot of users try to access the channel and the network is congested. Also, if strategy 3) is followed, some devices are not allowed to transmit in the considered frame and as a consequence the other nodes can strive for a larger fraction of the frame duration, thereby achieving lower distortion. A simple removal policy may exclude users to whom FOP$_\Gamma$ assigned the largest transmission times. Or, if different importance levels were assigned to different packets (or users), a priority-based policy could instead be used. %The former policy is more suitable for networks with homogeneous sensors, whereas the latter leads to better performance in heterogenous networks, e.g., in a monitoring system where the alarm sensor is given higher priority the the other devices. However, note that, if a user is dismissed, its data may be lost or useless, if a specific delivery deadline is missed.

 %TODO: citare le data queues nel system model?

\subsection{Notes on Convexity} \label{subsec:FOP_convex}

Solving the energy allocation problem in the next section would be much easier if $f_{\rm FOP}^{(k)}(\mathbf{E}^{(k)})$ were \emph{convex} in $\mathbf{E}^{(k)}$. In this subsection we prove the convexity for the particular case of low-SNR regime. While from our numerical evaluation this property seems to hold in general, a formal proof of this fact is left for future work. In any case, for the numerical results, we approximated $f_{\rm FOP}^{(k)}(\mathbf{E}^{(k)})$ with a convex function in order to correctly solve EAP.

\begin{thm} \label{thm:FOP_convex_E_lowSNR}
    In the low-SNR regime, FOP is convex in the input energy $\mathbf{E}^{(k)}$.
    \begin{proof}
        See \appendixname~\ref{proof:FOP_convex_E_lowSNR}.    
    \end{proof}
\end{thm}

\section{Energy-Allocation Problem (EAP)} \label{sec:EAP}

Network lifetime and average maximum distortion of the network are conflicting objectives (see~\eqref{eq:general_output}). EAP aims at finding the optimal energy allocation over time that balances these two quantities, according to the weight $\sigma$.

\subsection{Optimization Problem} \label{sec:EAP_optimization}

Assume the network lifetime $n$ is fixed. %(this is equivalent to fixing the weight $\sigma$). 
In this case, the optimization problem of Eq.~\eqref{eq:general_output} becomes~
\begin{subequations} \label{prob:EAP}
\begin{flalign} \label{eq:EAP_objective}
    \text{EAP:} && & \min_{\textstyle \mathbf{E} } \;\dfrac{1}{n} \sum_{k = 1}^n f_{\rm FOP}^{(k)}(\mathbf{E}^{(k)}), &
\end{flalign}
\vspace{-\belowdisplayskip}
\vspace{-\abovedisplayskip}
\begin{alignat}{2}
\shortintertext{subject to}
 & E_i^{(k)} \leq B_i^{(0)} - \sum_{j = 1}^{k-1} E_i^{(j)}, \quad && \forall i,\quad \forall k, \label{eq:EAP_batteries}\\
 & f_{\rm FOP}^{(k)}(\mathbf{E}^{(k)}) \mbox{ is feasible}, \quad && \forall k. \label{eq:EAP_feasible}
\end{alignat}
\end{subequations}

The objective function~\eqref{eq:EAP_objective} represents the average over $n$ frames of the maximum distortion achievable in every frame. It uses the function $f_{\rm FOP}^{(k)}(\cdot)$ as defined in Subsection~\ref{subsec:FOP_sol} and is convex according to Subsection~\ref{subsec:FOP_convex}. The size of the optimization variable $\mathbf{E}$ is known, as $n$ is fixed (see Eq.~\eqref{eq:general_output}). Eq.~\eqref{eq:EAP_batteries} is the energy causality constraint, which should be satisfied for every frame and for all users, and can be rewritten in equivalent but simpler form as~
\begin{align}
    \sum_{k = 1}^{n} E_i^{(k)} \leq B_i^{(0)}, \quad \forall i.
\end{align}

Finally, according to Subsection~\ref{subsec:FOP_unfeasible}, the last constraint imposes that FOP is feasible for every frame and for all users (see Section~\ref{subsec:FOP_unfeasible}). 
Note that the constraints induce a convex feasibility set because $f_{\rm FOP}^{(k)}(\mathbf{E}^{(k)})$ is convex in all the entries of $\mathbf{E}^{(k)}$.

In summary, EAP is a convex optimization problem and based on this observation we now propose a technique to solve it.
Consider matrix $\mathbf{E}$, and focus on the following problem, in which we optimize the sequence $E_\ell^{(1)},\ldots,E_\ell^{(n)}$ and keep all the other variables fixed:~
\begin{subequations} \label{prob:EAP_red}
\begin{flalign} \label{eq:EAP_red_objective}
    \text{EAP$_\ell$:} && & \min_{\textstyle \mathbf{E}_\ell} \sum_{k = 1}^n f_{\rm FOP}^{(k)}( \mathbf{E}^{(k)}), &
\end{flalign}
\vspace{-\belowdisplayskip}
\vspace{-\abovedisplayskip}
\begin{alignat}{2}
\shortintertext{subject to}
 & \sum_{k = 1}^{n} E_\ell^{(k)} \leq B_\ell^{(0)}, \label{eq:EAP_red_batteries}\\
 & \underline{E}_\ell^{(k)} \leq E_\ell^{(k)} \leq \overline{E}_\ell^{(k)}, \quad && \forall k. \label{eq:EAP_red_feasible}
\end{alignat}
\end{subequations}

\noindent $\underline{E}_\ell^{(k)}$ is defined as the minimum amount of energy that node $\ell$ should use in frame $k$ to obtain a feasible solution. Indeed, if $E_\ell^{(k)}$ were too low, it would not be possible to satisfy the distortion constraint~\eqref{eq:FOP_distortion} or the time constraint~\eqref{eq:FOP_tau} of FOP. Similarly, $\overline{E}_\ell^{(k)}$ is the energy value such that, for any $E_\ell^{(k)} \geq \overline{E}_\ell^{(k)}$, the objective function does not decrease further (i.e., after level $\overline{E}_\ell^{(k)}$, using more energy is useless). The values of $\underline{E}_\ell^{(k)}$ and $\overline{E}_\ell^{(k)}$ strictly depend on $\{E_1^{(k)},\ldots,E_{\ell-1}^{(k)},E_{\ell+1}^{(k)},\ldots,E_N^{(k)}\}$. As discussed in Subsection~\ref{subsec:FOP_convex}, the solution of FOP for frame $k$ is convex in $E_\ell^{(k)}$ and, in particular, it is strictly convex in $(\underline{E}_\ell^{(k)},\overline{E}_\ell^{(k)})$. Thus, EAP$_\ell$ can be solved in the dual domain by using the Lagrangian:~
\begin{subequations} \label{prob:EAP_red_L}
\begin{flalign} \label{eq:EAP_red_L_objective}
    && & \max_{\textstyle \lambda, \mathbf{E}_\ell} \;\sum_{k = 1}^n f_{\rm FOP}^{(k)}(\mathbf{E}^{(k)}) - \lambda \Big(\sum_{k = 1}^{n} E_\ell^{(k)} - B_\ell^{(0)} \Big)\!, &
\end{flalign}
\vspace{-\belowdisplayskip}
\vspace{-\abovedisplayskip}
\begin{alignat}{2}
\shortintertext{subject to}
 & \lambda \geq 0, \\
 & \underline{E}_\ell^{(k)} \leq E_\ell^{(k)} \leq \overline{E}_\ell^{(k)}, \quad && \forall k. \label{eq:EAP_red_L_feasible}
\end{alignat}
\end{subequations}

The Karush–Kuhn–Tucker conditions lead to~
\begin{align}
    &E_\ell^{(k)} = \max\{\underline{E}_\ell^{(k)},\min\{\overline{E}_\ell^{(k)},\theta^{-1}(\lambda)\} \}, \label{eq:E_ell_L}\\
    &\theta(E_\ell^{(k)}) \triangleq \frac{\partial f_{\rm FOP}^{(k)}(\mathbf{E}^{(k)})}{\partial E_\ell^{(k)}}, \label{eq:theta_derivative}
\end{align}

\noindent where $\lambda$ is such that $\sum_{k = 1}^{n} E_\ell^{(k)} = B_\ell^{(0)}$. %Eq.~\eqref{eq:E_ell_L} can be interpreted as a water-filling solution with minimum and maximum levels and in which the water level (i.e., the allocated energy) $\theta^{-1}(\lambda)$ is put in every slot $k$, if possible.

Since EAP$_\ell$ focuses on the optimization of one user at a time, we propose Algorithm~\ref{alg:AO} to solve the general problem.
Lines~\ref{alg:line:begin_AO}-\ref{alg:line:end_AO} perform the alternate optimization. In Line~\ref{alg:line:EAP_ell} we use matrix $\mathbf{E}$ to solve~\eqref{prob:EAP_red_L} and update its $\ell$-th row. Lines~\ref{alg:line:v_vect}-\ref{alg:line:end_AO} distribute in a random fashion the residual energy $B_\ell^{(0)}-\sum_{k = 1}^{n} E_\ell^{(k)}$ in all the slots where $E_\ell^{(k)}$ is equal to $\bar{E}^{(k)}$ ($\chi\{\cdot\}$ is the indicator function). Note that this operation does not change the distortion level obtained by solving EAP$_\ell$, but simply provides a new $\mathbf{E}_\ell$ that allows the alternate optimization to converge. 

We have the following result. 
 
\begin{algorithm}[!t]
\caption{Random Alternate Optimization}\label{alg:AO}
\begin{algorithmic}[1]
\State Initialize a feasible $\mathbf{E}$
\State $\overline{D} \gets \infty$
\While {$\overline{D}$ has not converged}
\For {$\ell = 1,\ldots,N$} \label{alg:line:begin_AO}
\State $\mathbf{E}_\ell \gets $ solve EAP$_\ell(\mathbf{E})$ \label{alg:line:EAP_ell}
\State $v \gets$ prob. vector of size $\sum_k \chi\{E_\ell^{(k)} = \overline{E}^{(k)}\}$ \label{alg:line:v_vect}
\State $S \gets \sum_{k = 1}^{n} E_\ell^{(k)}$ \label{alg:line:S}
\State $v_{\rm ind} \gets 1$
\For {$k = 1,\ldots,n$ \textbf{ such that } $E_\ell^{(k)} = \overline{E}^{(k)}$}
\State $E_\ell^{(k)} \gets v(v_{\rm ind})\cdot (B_\ell^{(0)}-S)$
\State $v_{\rm ind} \gets v_{\rm ind} + 1$  \label{alg:line:end_AO}
\EndFor
\EndFor
\State $\overline{D} \gets 1/n \;\sum_{k = 1}^n f_{\rm FOP}^{(k)}(\mathbf{E}^{(k)})$
\EndWhile
\end{algorithmic}
\end{algorithm}

\begin{thm}\label{thm:alternate_opt}
    The alternate optimization approach of Algorithm~\ref{alg:AO}, in which only user $\ell$ is considered in a single step, leads to the optimal solution.
    \begin{proof}
        See \appendixname~\ref{proof:alternate_opt}.
    \end{proof}
\end{thm}

In summary, EAP solves the energy allocation problem over time using an alternate optimization procedure. At every step of the algorithm (Line~\ref{alg:line:EAP_ell}), EAP$_\ell$ is solved, and FOP is invoked multiple times to evaluate the derivative in~\eqref{eq:theta_derivative}, which relates the allocated energy to the corresponding distortion metric.
This is iterated over all possible values of $n$ and, by tuning $\sigma$, the network designer can choose a point in the trade-off between lifetime and QoS.

%Let $\bar{D}$ be the average distortion value (see~\eqref{eq:FOP_minmax}) obtained by EAP for a fixed lifetime $n$ by using the alternate optimization approach.

\section{Numerical Evaluation}\label{sec:numerical_evaluation}

In this section we show how the system parameters influence the distortion of the system. We consider five groups of nodes with different distortion curves placed at a fixed distance $d$ from the BS.

If not otherwise stated, we use the following parameters. The frame duration $T$ is $0.15$~ms. The channel gains are computed using the standard path loss model with a path-loss exponent equal to $3.5$ (e.g., as in an urban scenario) and a central frequency $915$~MHz. The bandwidth is $W = 125$~kHz, and the overall noise power is $-167$~dBm. The parameters of the distortion curves are $\alpha_i \in [0.35,0.63,0.69,0.57,0.82]$ and $b_i = [19.9,3.44,3.27,9.94,6.35]$, which have been derived through empirical fittings of the realistic rate-distortion curves of~\cite{zordan}. The energy consumption model assumes $\varepsilon_{{\rm C},i} = 5\cdot 10^{-7}$~W, $\beta = 10^{-4}$~J for all frames, and an initial battery level $B_{0,i} = 5$~mJ. The minimum and maximum power consumptions are $P_{{\rm min},i} = 0$~W and $P_{{\rm max},i} = 25$~mW. Finally, we impose a distortion threshold $D_{\rm th} = 8\%$ for packets with fixed size $L_{0,i} = 500$~bits. These values are not associated to any specific protocol, but are reasonable and suitable for WSNs. 
%We do not fix weight $\sigma$, whose value dictates the working point of the tradeoff between lifetime and distortion.

\begin{figure}[t]
  \centering
  \includegraphics[trim = 0mm 0mm 0mm 0mm,  clip, width=1\columnwidth]{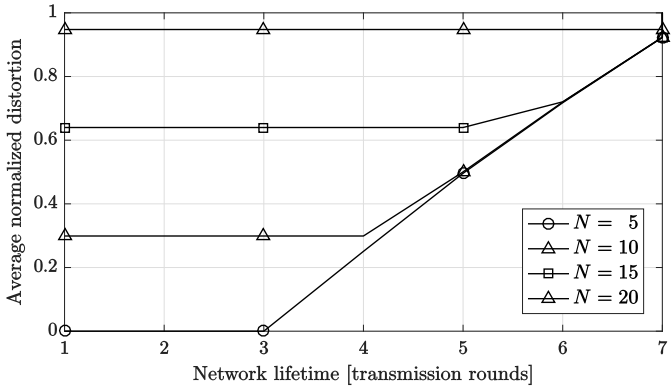}
  \caption{Optimal normalized distortion as a function of the lifetime $n$ for different number of nodes when $d = 250$~m.}
  \label{fig:D_change_N}
\end{figure}

In \figurename~\ref{fig:D_change_N} we plot the distortion and the lifetime obtained as solutions of the optimization problem~\eqref{eq:general_output} for different values of the number of nodes $N$, the curves have been obtained by changing the weight factor $\sigma$. The distortion tends to increase with the lifetime, as expected, since smaller amounts of energy can be allocated in each frame and thus nodes must compress more to transmit their data. For small values of $n$, the graphs are constant because not all the energy in the batteries is used or because a zero distortion is achieved. Clearly, it is always better to choose the right extremes of the constant regions rather than the other points since they provide the same QoS with longer lifetimes. The maximum lifetime is reached when the problem becomes infeasible (see condition i2) of Section~\ref{subsec:FOP_unfeasible}), i.e., the curves in \figurename~\ref{fig:D_change_N} reach value $1$ and no energy allocation can satisfy all constraints of FOP with an acceptable distortion for all the frames of the considered lifetime.
We also remark that the lifetime strongly depends on the initial battery level. However, even if larger batteries were considered, the trend of the distortion curves would remain the same. Note that, because of the symmetry of our setup, all curves coincide after a certain value of $n$. This happens because the only way to reach high values of $n$ is to assign low energy to every frame, which in turn corresponds to short $\tau_i^{(k)}$ and thus Constraint~\eqref{eq:FOP_tau} is always satisfied. Future work includes the study of more complex scenarios using different node locations and Montecarlo simulations.

\begin{figure}[t]
  \centering
  \includegraphics[trim = 0mm 0mm 0mm 0mm,  clip, width=1\columnwidth]{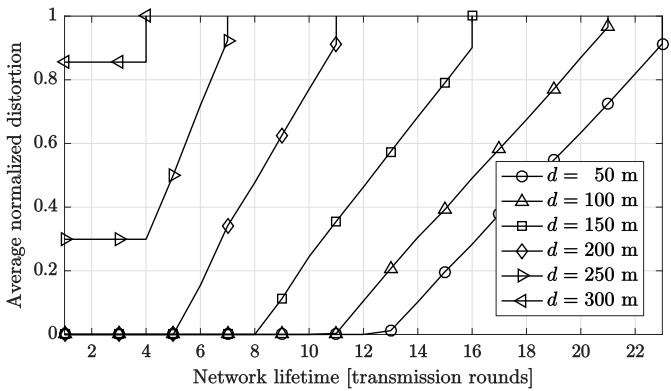}
  \caption{Optimal normalized distortion as a function of the lifetime $n$ for different distances when $N = 10$.}
  \label{fig:D_change_distance}
\end{figure}
In \figurename~\ref{fig:D_change_distance}, we show how the distance $d$ (equal for all nodes) influences the network performance. 
As $d$ increases, the performance of the system clearly decays and, when $d$ is extremely high, the network cannot operate at all. It is interesting to note that, because of the strong path loss, the higher $d$, the higher the gap between two adjacent curves.

\begin{figure}[t]
  \centering
  \includegraphics[trim = 0mm 0mm 0mm 0mm,  clip, width=1\columnwidth]{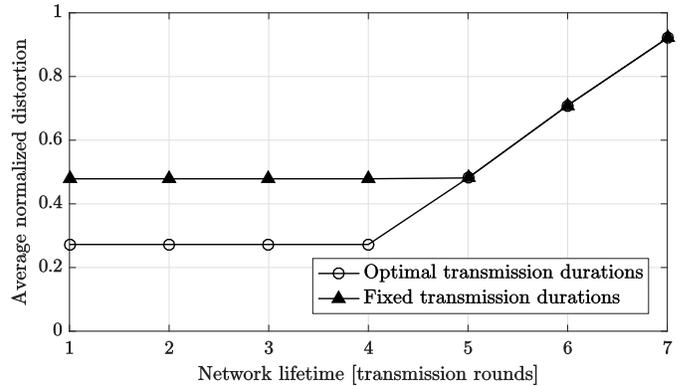}
  \caption{Normalized distortion evaluated optimally and with fixed transmission durations as a function of the lifetime $n$ when $N = 10$ and $d = 250$~m.}
  \label{fig:D_fixed}
\end{figure}

Finally, \figurename~\ref{fig:D_fixed} shows the comparison between the optimal approach of Eq.~\eqref{eq:general_output} which evaluates the transmission durations according to FOP, and a sub-optimal approach which uses fixed durations (i.e., $\tau_i^{(k)} = T/N,\ \forall i,k$). Note that all nodes are located at the same distance, thus there is no near-far effect (which however would further support our approach). Even in this case, because of the different rate-distortion curves, the optimal solution may provide much better performance than the naive scheme, which confirms the importance of adapting the protocol parameters to the characteristics of the data sources in a real deployment.
In all cases we analyzed, the distortion levels stay below the tolerable threshold, i.e., the sink can recover all data with pre-defined accuracy. When the normalized distortion is 0, the information gathered from the sensors is sent over the channel with the highest possible quality (i.e., no loss), whereas a normalized distortion close to 1 corresponds to a very rough quantization of the signals being sent. The correspondence between the actual values of the normalized distortion and the quality perceived by the user will depend on the specific application, and a detailed study of this relationship will be left as future work.

\section{Conclusions}\label{sec:conclusions} 
In this paper, we studied a TDMA-based strategy for battery-powered devices with QoS requirements.
We defined an optimal scheduling policy that jointly considers source coding operations and energy constraints, by determining the data compression ratio and the energy allocation for each node in each time frame, respectively. In particular, using convex and alternate programming, we presented an efficient method to minimize the average of the maximum distortions for a given lifetime. Numerical results show the importance of using optimized protocols to compensate the near-far effect.

Future work includes more detailed numerical results, the study of latency on the data transmission, and the investigation of online schemes to derive the optimal performance.

\section*{Acknowledgments}
This work was partially supported by Intel's Corporate Research Council.

\appendices

\section{Proof of Lemma~\ref{lemma:FOP_gamma}}\label{proof:FOP_gamma}

By contradiction, assume that there exists $\ell \in \mathcal{N}$ such that a distortion $D_\ell^\prime < \Gamma^\star D_{{\rm th},\ell}$ leads to an optimal solution, but $D_\ell^{\prime\prime} = \Gamma^\star D_{{\rm th},\ell}$ does not. According to~\eqref{eq:dist}, this implies that longer (higher quality) packets are used in the first case, i.e., $L_\ell^\prime > L_\ell^{\prime \prime}$. Then, if we chose $P_{{\rm tx},\ell}^{\prime\prime} = P_{{\rm tx},\ell}^\prime$ and $\tau_\ell^{\prime\prime} = \tau_\ell^\prime L_\ell^{\prime\prime} / L_\ell^\prime$, all constraints~\eqref{eq:FOP_distortion}-\eqref{eq:FOP_tau} would still be satisfied because $\tau_\ell^{\prime\prime} < \tau_\ell^\prime$. Thus we would find a feasible optimal solution in which~\eqref{eq:FOP_gamma} is satisfied with equality, i.e., the initial assumption must be wrong.
        
\section{Proof of Theorem~\ref{thm:Gamma1_Gamma2}}\label{proof:Gamma1_Gamma2}

Consider a generic node $\ell$. According to~\eqref{eq:P_TX_star} and thanks to Assumption~\ref{ass:P_max_high}, the point $P_{{\rm tx},\ell}^\star$ always falls in the increasing right branch of $g_\ell(x)$ (otherwise it is equal to $P_{{\rm max},\ell}$).
        
By assumption we have $\Gamma^\prime \leq \Gamma^{\prime\prime}$, which, using Definition~\eqref{eq:dist}, implies $L_\ell^\prime \geq L_\ell^{\prime\prime}$. The right-hand side of~\eqref{eq:P_TX_E} is a decreasing function of $L_\ell$, thus, by naming $P^\prime$ and $P^{\prime\prime}$ the points $P_{{\rm tx},i}^{E_i}$ corresponding to $\Gamma^\prime$ and $\Gamma^{\prime\prime}$, respectively, we obtain $P^\prime \leq P^{\prime\prime}$. Also, since $P^\prime \geq P_{{\rm min},\ell}$ exists by assumption, also $P^{\prime\prime}$ exists.
        
Finally, since the same energy $E_\ell$ is considered in the two cases, $P^\prime$ leads to a transmission duration $\tau_\ell^\prime$ longer than what can be found with $P^{\prime\prime}$. Thus, since with $P^\prime$ all the constraints of Problem~\eqref{prob:FOP} were satisfied, also using $P^{\prime\prime}$ they remain satisfied (in particular, Constraint~\eqref{eq:FOP_tau} is still satisfied because $\tau_\ell^{\prime\prime} \leq \tau_\ell^\prime$).

\section{Proof of Theorem~\ref{thm:FOP_convex_E_lowSNR}}\label{proof:FOP_convex_E_lowSNR}

Under the assumption of low-SNR regime, we have:~
\begin{align}
    L_i \simeq  \tau_i \,W\, h_i\,P_{{\rm tx},i}\log_2 {\rm e}, \quad \forall i. \label{eq:FOP_L_low}
\end{align}

By combining it with constraint~\eqref{eq:FOP_energy_const}, we obtain that the size $L_i$ of the transmitted packet is linear in the given energy $E_i^{(k)}$ for each user. Then, according to~\eqref{eq:FOP_distortion}, and taking into account that $\alpha_i > 0 \:\forall i$, $D_i$ is convex in $E_i^{(k)}, \:\forall i$. Finally, since the maximum operation preserves convexity, $f_{\rm FOP}^{(k)}(\mathbf{E}^{(k)})$ is convex.

\section{Proof of Theorem~\ref{thm:alternate_opt}}\label{proof:alternate_opt}

The optimization problem is convex, but non-strictly convex in general. Thus a unique minimum, namely $\Psi^\star$, exists but, potentially, with multiple minimum points.         
We now show that Algorithm~\ref{alg:AO} produces a non-increasing sequence of distortion values ($\Psi_1 \geq \Psi_2 \geq\ldots$), and, unless the minimum is achieved, there exists a non-null probability that $\Psi_m > \Psi_{m+1}$, for some $m$. Since the minimum is unique, the sequence converges to $\Psi^\star$ thanks to~\cite[Proposition~2.7.1]{bertsekas}.

        %While in general the order TODO Gauss–Seidel vs arbitrary 
        %Gauss–Seidel type, which minimizes F cyclically over each of x1, . . . , xs while fixing the remaining blocks at their last updated values
        
Consider the generic step $m$ of the algorithm, in which the optimization revolves around node $\ell$, i.e.,  we focus the sequence $\mathbf{E}_\ell$. As described in Section~\ref{subsec:FOP_convex}, the objective $\Gamma_\ell^{(k)}$ is a convex function of $E_\ell^{(k)}$. In particular, $\Gamma_\ell^{(k)}$ is strictly convex in $[\underline{E}_\ell^{(k)},\overline{E}_\ell^{(k)}]$, and constant for $E_\ell^{(k)} \geq \overline{E}_\ell^{(k)}$. Thus, we can simplify the problem by restricting our attention to the strictly decreasing region, and handle the constant region subsequently.

Since the sum of convex functions is convex, when we aim at optimizing $\sum_{k = 1}^n \Gamma_\ell^{(k)}$ in $[\underline{\mathbf{E}}_\ell,\overline{\mathbf{E}}_\ell]$, this is a convex optimization problem and can be solved using~\eqref{prob:EAP_red_L}. Let $\mathbf{E}_\ell^\star$ be the solution. Two cases should be considered:
\begin{enumerate}
    \item All the elements of $\mathbf{E}_\ell^\star$ fall inside the strictly convex region $[\underline{\mathbf{E}}_\ell,\overline{\mathbf{E}}_\ell)$ ($\overline{\mathbf{E}}_\ell$ is excluded). This may happen because of the battery constraints. In this case, there is only one optimal solution, and no other actions are required for the current step ($S = 0$ in Line~\ref{alg:line:S} of the algorithm);
    \item Some of the elements fall at the beginning of the constant region, i.e., $E_\ell^{(k)} = \overline{E}_\ell^{(k)}$ for some $k \in \overline{\mathcal{K}}$. In this case, also other solutions may be optimal for the current iteration, since using all the feasible energy combinations with $E_\ell^{(k)} \succeq \overline{E}_\ell^{(k)}$ for $k \in \overline{\mathcal{K}}$ lead to the same solution. However, although all these sequences provide the same $\Psi_m$ in the current step, they may influence the future values $\Psi_{m+1},\Psi_{m+2},$ etc. Then, two subcases should be distinguished: in the following $N-1$ steps of the algorithm (i.e., the next time that node $\ell$ is examined), the sequence of $\Psi$ has either strictly decreased, or remained constant. In the former case, the algorithm proceeds toward the optimal solution. In the latter, the algorithm cyclically returns to point $m$, thus the sequence of $\Psi$ has not improved. In this case, we choose other points $E_\ell^{(k)} > \overline{E}_\ell^{(k)},\, k \in \overline{\mathcal{K}}$ (e.g., with a random approach as described in Lines~\ref{alg:line:v_vect}-\ref{alg:line:end_AO}) and repeat the procedure. In an infinite horizon, all the possible energy combinations have been tested with a non-null probability, thus the algorithm has proceeded toward an optimal solution.
\end{enumerate}

In the worst case scenario, Algorithm~\ref{alg:AO} may degenerate in almost an exhaustive search; however, in practical cases, very few iterations are required, and the algorithm rapidly converges.

\bibliographystyle{IEEEtran}
\bibliography{bib}

\end{document}